\pdfoutput=1
\documentclass[11pt, a4paper]{article}
\usepackage[utf8]{inputenc}
\usepackage{amsmath, amssymb, amsfonts, bm, cite, float}
\usepackage{geometry}

\begin{document}

\begin{titlepage}
\vspace{2cm}
\begin{center}
{\Large\bf Supersymmetric Origin of Four-Dimensional Space-time \\ in the IIB Matrix Model}
\vspace{2cm}

{\large Tetsuyuki Muramatsu\footnote{The views expressed in this paper are solely those of the author and do not necessarily reflect those of the affiliated organization.}} \\

\vspace{1cm}
{\it Tokai National Higher Education and Research System \\
Furo-cho, Chikusa-ku, Nagoya, 464-8601, Japan} \\
\vspace{2cm}

{\bf Abstract}
\end{center}
We investigate the constraints imposed by supersymmetry on the IIB matrix model (IKKT model) by requiring both the closure of the transformations and the satisfaction of the Ward identities at the leading order of the order expansion. Following the systematic methodology, we evaluate the most general forms of the effective action and supersymmetry transformations consistent with the $SU(N)$ algebra. 
In ten dimensions, we prove that these supersymmetric requirements lead to a non-renormalization theorem, which forces all coefficient functions to be constant. This result stems from the emergence of a 5-form tensor in the closure condition that cannot be absorbed by the $SU(N)$ algebra. This residual term strictly forbids non-trivial fluctuations at the leading order.
While a similar non-renormalization theorem holds in four dimensions, we demonstrate that the four-dimensional Clifford algebra provides a unique exit through Hodge duality. 
This duality geometrically maps the anomalous high-rank tensor structures into absorbable lower-rank forms, allowing for non-trivial dynamical backgrounds prohibited in ten dimensions.
We find that such non-trivial solutions are restricted to (anti-)self-dual configurations, which, through reality conditions, necessitate a Euclidean metric. Our results indicate that the emergence of a four-dimensional Euclidean space-time is a prerequisite for the theory to admit non-trivial backgrounds while preserving supersymmetry at the leading order.

\end{titlepage}

\newpage

\section{Introduction}

The IIB matrix model, also known as the IKKT model \cite{Ishibashi:1996xs}, was proposed as a promising non-perturbative formulation of superstring theory. One of its most fascinating aspects is its potential to dynamically generate the four-dimensional spacetime of our universe from its fundamental ten-dimensional matrix degrees of freedom via spontaneous symmetry breaking (SSB). 

The mechanism of this dimensional reduction has been extensively investigated primarily through dynamical approaches, such as Monte Carlo simulations of the Lorentzian matrix model \cite{Kim:2011cr, Ito:2014, Nishimura:2019, Anagnostopoulos:2023, Anagnostopoulos:2026} and evaluations of the effective action. 
In this paper, complementing these previous studies, we aim to clarify how four dimensions can emerge by focusing on the kinematical constraints imposed by supersymmetry on the effective action.

Historically, the constraints imposed by extended supersymmetry on the effective action of matrix quantum mechanics have been recognized as highly restrictive \cite{Paban:1998ea}. Building upon this concept, we investigate the stringent requirements imposed by the off-shell closure of the supersymmetry algebra and the satisfaction of the Ward identities\cite{Kazama:2002cb}. To systematically analyze the low-energy effective theory, we employ the order expansion scheme. By strictly adhering to the $SU(N)$ algebra---which inherently excludes the $U(1)$ center-of-mass degrees of freedom---we construct the most general forms of the effective action and the supersymmetry transformations at the leading order.

Through this algebraic approach, we first evaluate the ten-dimensional case. We prove that requiring the closure of the supersymmetry transformations inevitably yields a residual 5-form tensor structure. Because the $SU(N)$ algebra lacks any corresponding generator to absorb this 5-form, the theory is forced into a strict non-renormalization theorem where all coefficient functions must be constant. This implies that, at the leading order, non-trivial dynamical fluctuations are strictly forbidden in ten dimensions.

We then demonstrate that a similar non-renormalization theorem is also established upon reduction to four dimensions. However, to explore the possibility of retaining non-trivial dynamics within these constraints, we investigate (anti-)self-dual background configurations. 
Remarkably, the four-dimensional Clifford algebra, through the mechanism of Hodge duality, reduces the decomposed high-rank tensor obstructions into absorbable lower-rank forms.
This provides a potential ``algebraic exit'' that might allow the algebra to close non-trivially. We find that such dynamically fluctuating solutions are indeed restricted to (anti-)self-dual configurations, which, due to the reality conditions for Hermitian matrices, can only be realized if the metric is Euclidean \cite{Nishimura:2001sx}. Therefore, within the scope of our leading-order analysis, the emergence of a four-dimensional Euclidean spacetime represents a compelling possibility for accommodating non-trivial backgrounds while preserving supersymmetry, though we acknowledge that this conclusion might be subject to modifications when higher-order interactions are taken into account.

This paper is organized as follows. In Section 2, we define our notation, the $SU(N)$ framework, and the order expansion scheme. In Section 3, we detail the evaluation of the algebraic closure in ten dimensions and establish the non-renormalization theorem. In Section 4, we establish the four-dimensional non-renormalization theorem and elucidate the potential rescue mechanism provided by Hodge duality. Section 5 discusses the constraints that necessitate the Euclidean metric and the (anti-)self-dual configurations. Finally, in Section 6, we conclude our findings and briefly discuss the limitations of our leading-order analysis alongside potential directions for future higher-order investigations.

\section{General Formulations and the Weighting Scheme}

In this section, we define the theoretical framework of our analysis. We first introduce the original IIB matrix model and its underlying supersymmetry. We then introduce a systematic weighting scheme, which clarifies the expansion structure of the effective action, the supersymmetry transformations, and the associated Ward identities.

\subsection{The IIB Matrix Model and Original Supersymmetry}
We start with the original action of the IIB matrix model (IKKT model) with the gauge group $SU(N)$. The basic variables are ten $N \times N$ Hermitian matrices $X_\mu$ ($\mu = 0, \dots, 9$) and ten-dimensional Majorana-Weyl spinors $\Psi$. The action is given by
\begin{equation}
    S = -\frac{1}{g^2} \mathrm{Tr} \left( \frac{1}{4} [X_\mu, X_\nu][X^\mu, X^\nu] + \frac{1}{2} \bar{\Psi} \Gamma^\mu [X_\mu, \Psi] \right) ,
\end{equation}
where $g$ is the coupling constant, and the spacetime indices are contracted using the flat metric $\eta_{\mu\nu} = \mathrm{diag}(-1, 1, \dots, 1)$. In the $SU(N)$ formulation, we strictly exclude the $U(1)$ sector, meaning $X_\mu$ and $\Psi$ are traceless matrices.

Crucially, this original action is invariant under the following classical supersymmetry transformations:
\begin{align}
    \delta X_\mu &= i\bar{\epsilon} \Gamma_\mu \Psi , \\
    \delta \Psi &= \frac{i}{2} [X_\mu, X_\nu] \Gamma^{\mu\nu} \epsilon ,
\end{align}
where $\epsilon$ is a constant Majorana-Weyl spinor parameter. 

To study the low-energy dynamics, we decompose the matrices into a classical background and quantum fluctuations. By integrating out the fluctuations in the path integral, we obtain the low-energy effective action $\Gamma$. In what follows, we denote the background bosonic fields as $B_\mu$ and the fermionic fields as $\lambda$. Two fundamental quantities characterize our formulation. First, the radial coordinate $r$, which serves as the fundamental scale of the background configuration, is defined as
\begin{equation}
    r = \sqrt{\frac{1}{N} \mathrm{Tr}(B_\mu B^\mu)} . \label{eq:def_r}\footnote{As noted in the definition of the action, the indices are contracted using the Minkowski metric. To ensure that the differential operator $\partial_r$ is mathematically well-defined in this section, we implicitly focus on configurations where $r^2 > 0$. This signature-related subtlety is naturally resolved by the transition to the Euclidean metric in Section 5.}
\end{equation}
Second, the field strength, characterizing the non-trivial dynamical fluctuations, is defined by the commutator
\begin{equation}
    \mathcal{F}_{\mu\nu} = i[B_\mu, B_\nu] . \label{eq:def_F}
\end{equation}

\subsection{The Weighting Scheme and the Expansion Structure}
Following the methodology established in \cite{Kazama:2002cb}, we introduce a systematic weighting scheme to expand the effective action. We assign a specific order to the operators:
\begin{itemize}
    \item Each commutator (or adjoint action) $[B_\mu, \cdot]$ is assigned order 1.
    \item Each fermion bilinear $(\bar{\lambda} \dots \lambda)$ is assigned order 1.
    \item The background field $B_\mu$ itself is assigned order 0.
\end{itemize}
Because of Lorentz invariance and the trace structure, the effective action $\Gamma$ consists strictly of even-order terms: 
\begin{equation}
    \Gamma = \sum_{m=1}^{\infty} \Gamma^{(2m)} \quad (2m = 2, 4, 6, \dots) .
\end{equation}

\subsection{General Form of the Effective Theory and Discrete Symmetries}
To construct the most general forms of the effective action $\Gamma^{(2)}$ and the modified supersymmetry transformations $\delta_\epsilon^{(1)}$ at the leading order, we heavily rely on Lorentz invariance and the discrete symmetries of the original 10-dimensional Super Yang-Mills theory, specifically C-symmetry (charge conjugation) and CPT symmetry. 

Moreover, the number of independent terms is drastically reduced by utilizing the algebraic properties of 10-dimensional Majorana-Weyl spinors. For anti-commuting spinors $\lambda$ and $\chi$, the bilinears satisfy the following fundamental symmetry property:
\begin{equation}
    \bar{\lambda} \Gamma_{\mu_1 \dots \mu_n} \chi = -(-1)^{\frac{n(n-1)}{2}} \bar{\chi} \Gamma_{\mu_1 \dots \mu_n} \lambda . \label{eq:bilinear_sym}
\end{equation}
Combined with the cyclic property of the trace over color indices, many terms naturally vanish. Furthermore, any redundant fermion structures are absorbed and simplified by repeatedly applying the Fierz identity, whose representative basic form is
\begin{equation}
    \epsilon_1 \bar{\epsilon}_2 = \frac{1}{32} \sum_{n} \frac{1}{n!} (\bar{\epsilon}_2 \Gamma_{\mu_1 \dots \mu_n} \epsilon_1) \Gamma^{\mu_1 \dots \mu_n} . \label{eq:fierz_base}
\end{equation}

Applying these strict symmetrical and algebraic constraints, the most general effective action up to order 2 is narrowed down to
\begin{equation}
    \Gamma^{(2)} = f(r) \mathrm{Tr}(\mathcal{F}_{\mu\nu}\mathcal{F}^{\mu\nu}) + g(r) \mathrm{Tr}(\bar{\lambda} \Gamma^\mu [B_\mu, \lambda]) + h(r) \mathrm{Tr}(\bar{\lambda} \Gamma_{\mu\nu\rho} \lambda \bar{\lambda} \Gamma^{\mu\nu\rho} \lambda) . \label{eq:Gamma2}
\end{equation}
Similarly, the most general supersymmetry transformation at order 1 (which increases the order of operators by 1) is given by
\begin{align}
    \delta_\epsilon^{(1)} B_\mu &= c(r) \bar{\epsilon} \Gamma_\mu \lambda , \label{eq:delta1_B} \\
    \delta_\epsilon^{(1)} \lambda &= w(r) \mathcal{F}_{\mu\nu} \Gamma^{\mu\nu} \epsilon + u_1(r) (\bar{\lambda} \Gamma_\rho \lambda) \Gamma^\rho \epsilon + u_2(r) (\bar{\lambda} \Gamma_{\rho\sigma\tau} \lambda) \Gamma^{\rho\sigma\tau} \epsilon , \label{eq:delta1_L}
\end{align}
where the coefficients $f, g, h, c, w, u_i$ are functions of $r$ of order 0.

\subsection{Ward Identities and Algebraic Closure}
The effective theory must satisfy the Ward identity reflecting the underlying supersymmetry, $\delta_\epsilon \Gamma = 0$. Abstractly, expanding this algebraically yields a hierarchy of constraints at each total order $n$:
\begin{equation}
    \sum_{k=1}^{n-2} \delta_\epsilon^{(k)} \Gamma^{(n-k)} = 0 . \label{eq:ward_hierarchy}
\end{equation}
Requiring this Ward identity at the lowest non-trivial order (total order $n=3$, yielding $\delta_\epsilon^{(1)} \Gamma^{(2)} = 0$) imposes severe restrictions on the coefficient functions.

Simultaneously, the modified transformations must satisfy the algebraic closure. The commutator of two transformations must abstractly close onto the $SU(N)$ algebra:
\begin{align}
    [\delta_{\epsilon_1}^{(1)}, \delta_{\epsilon_2}^{(1)}] B_\mu &= \delta_{\mathrm{gauge}}(\Lambda) B_\mu + \delta_{\mathrm{Lorentz}}(\omega) B_\mu + \delta_{\mathrm{EoM}} B_\mu , \label{eq:closure_B} \\
    [\delta_{\epsilon_1}^{(1)}, \delta_{\epsilon_2}^{(1)}] \lambda &= \delta_{\mathrm{gauge}}(\Lambda) \lambda + \delta_{\mathrm{Lorentz}}(\omega) \lambda + \delta_{\mathrm{EoM}} \lambda . \label{eq:closure_L}
\end{align}
The strict exclusion of the $U(1)$ sector absolutely precludes any translation or central charge terms on the right-hand side. The detailed evaluation of this commutator, which reveals the critical properties of the ten-dimensional theory, will be thoroughly executed in the next section.

\section{Evaluation of the Algebraic Closure in Ten Dimensions}

In this section, we explicitly evaluate the closure of the supersymmetry algebra in ten dimensions. By analyzing the commutator of the transformations on the bosonic field, we demonstrate the inevitable emergence of a 5-form tensor structure, which directly leads to the exact non-renormalization theorem.

\subsection{The Commutator on the Bosonic Field}
Let us evaluate the commutator $[\delta_{\epsilon_1}^{(1)}, \delta_{\epsilon_2}^{(1)}]$ acting on the bosonic field $B_\mu$. Recalling the transformation $\delta_\epsilon^{(1)} B_\mu = c(r) \bar{\epsilon} \Gamma_\mu \lambda$, the commutator naturally splits into two distinct parts:
\begin{equation}
    [\delta_{\epsilon_1}^{(1)}, \delta_{\epsilon_2}^{(1)}] B_\mu = c(r) \bar{\epsilon}_2 \Gamma_\mu (\delta_{\epsilon_1}^{(1)} \lambda) + \delta_{\epsilon_1}^{(1)}(c(r)) \bar{\epsilon}_2 \Gamma_\mu \lambda - (\epsilon_1 \leftrightarrow \epsilon_2) .
\end{equation}
The first term, involving the variation of the field $\lambda$, contributes to the standard gauge, Lorentz, and equations of motion (EoM) terms. Our primary focus is the second term, which involves the variation of the coefficient function $c(r)$.

\subsection{Emergence of the Derivative Terms}
The variation of $c(r)$ originates from the variation of the radial coordinate $r$. Using Eq.~\eqref{eq:def_r}, its variation under the supersymmetry transformation is given by
\begin{equation}
    \delta_\epsilon^{(1)} r = \frac{1}{2rN} \mathrm{Tr}(\delta_\epsilon^{(1)} B_\nu B^\nu + B_\nu \delta_\epsilon^{(1)} B^\nu) = \frac{c(r)}{rN} \mathrm{Tr}(B^\nu \bar{\epsilon} \Gamma_\nu \lambda) .
\end{equation}
Substituting this into the commutator, we obtain anomalous terms proportional to the derivative $c'(r)$:
\begin{equation}
    \Delta B_\mu^{(deriv)} = \frac{c(r)c'(r)}{rN} \left[ \mathrm{Tr}(B^\nu \bar{\epsilon}_1 \Gamma_\nu \lambda) \bar{\epsilon}_2 \Gamma_\mu \lambda - (\epsilon_1 \leftrightarrow \epsilon_2) \right] . \label{eq:deltaB_deriv}
\end{equation}

\subsection{Fierz Transformation and the 5-Form Obstruction}
To verify the consistency of the closure condition, we must isolate the transformation parameters $\epsilon_{1,2}$ from the dynamical fields. We define the anti-symmetric bilinear parameters:
\begin{equation}
    v^{(n)}_{\mu_1 \dots \mu_n} = \bar{\epsilon}_2 \Gamma_{\mu_1 \dots \mu_n} \epsilon_1 .
\end{equation}
In ten dimensions, for Majorana-Weyl spinors, the only non-vanishing anti-symmetric parameters are the 1-form $v^{(1)}_\rho$ and the self-dual 5-form $v^{(5)}_{\rho\sigma\tau\lambda\eta}$. To explicitly see how the 5-form emerges, we utilize the Fierz identity for the anti-symmetrized product of two transformation parameters:
\begin{equation}
    \epsilon_1 \bar{\epsilon}_2 - \epsilon_2 \bar{\epsilon}_1 = \frac{1}{16} v^{(1)}_\rho \Gamma^\rho + \frac{1}{3840} v^{(5)}_{\rho\sigma\tau\lambda\eta} \Gamma^{\rho\sigma\tau\lambda\eta} . \label{eq:fierz_obstruction}
\end{equation}
By substituting this expansion into Eq.~\eqref{eq:deltaB_deriv}, the derivative contribution is decomposed into the 1-form and 5-form sectors. Most critically, this rearrangement inevitably exposes a term proportional to the 5-form parameter:
\begin{equation}
    \Delta B_\mu^{(deriv)} \supset \frac{c(r)c'(r)}{rN} v^{(5)}_{\rho\sigma\tau\lambda\eta} \mathrm{Tr} \left( B^\nu \bar{\lambda} \Gamma_\nu \Gamma^{\rho\sigma\tau\lambda\eta} \Gamma_\mu \lambda \right) .
\end{equation}
This demonstrates the structural existence of the 5-form parameter within the closure condition.

\subsection{The 10D Non-Renormalization Theorem}
We now confront this result with the fundamental requirement of algebraic closure. The right-hand side of the closure condition \eqref{eq:closure_B} must consist strictly of generators of the $SU(N)$ algebra. However, the $SU(N)$ algebra possesses absolutely no generator capable of absorbing the 5-form parameter $v^{(5)}$. Furthermore, the 1-form parameter $v^{(1)}$, corresponding to translations, is also precluded as the $U(1)$ sector is decoupled.

Consequently, these terms must vanish identically, imposing the strict condition $c(r)c'(r) = 0$. To avoid a trivial transformation, we are forced to conclude
\begin{equation}
    c'(r) = 0 , \quad \implies \quad c(r) = c_0 (\text{const.}) .
\end{equation}
This single constraint triggers a rigid cascade of algebraic relations through the remaining requirements of supersymmetry:
\begin{enumerate}
    \item \textbf{Closure on $\lambda$:} The requirement that Eq.~\eqref{eq:closure_L} reproduces $SU(N)$ generators fixes the products of the coefficients to specific numerical constants $\alpha, \beta_i$:
    \begin{equation}
        c(r) w(r) = \alpha , \quad c(r) u_i(r) = \beta_i . \label{eq:lock_trans}
    \end{equation}
    \item \textbf{Ward Identity:} The requirement $\delta_\epsilon^{(1)} \Gamma^{(2)} = 0$ enforces the cancellation of variations between the functions $f, g, h$ of the effective action:
    \begin{equation}
        f(r) c(r) = \kappa_1 g(r) w(r) , \quad h(r) c(r) = \kappa_2 g(r) u_i(r) , \label{eq:lock_action}
    \end{equation}
    where $\kappa_1, \kappa_2$ are numerical factors.
\end{enumerate}
Since $c(r) = c_0$, Eqs.~\eqref{eq:lock_trans} and \eqref{eq:lock_action} demand that $w(r)$, $u_i(r)$, $f(r)/g(r)$, and $h(r)/g(r)$ are all pure constants. Ultimately, the effective action $\Gamma^{(2)}$ is forced to be precisely proportional to the original action of the IIB matrix model. This establishes the non-renormalization theorem in ten dimensions: the dynamical fluctuations are algebraically frozen.

\section{The Four-Dimensional Non-Renormalization Theorem and the Algebraic Exit}

In the previous section, we established the non-renormalization theorem in ten dimensions. In this section, we investigate the four-dimensional dimensional reduction. By explicitly tracking the coefficients in the commutator and utilizing the inherent Weyl nature of the 4-dimensional spinors, we mathematically demonstrate exactly why the generalized self-duality condition is required, and how it poses a fundamental dilemma in Minkowski spacetime.

\subsection{$SO(1,3) \times SO(6)$ Decomposition and Internal Spinors}

We decompose the original ten-dimensional Minkowski background into four
spacetime gauge fields $B_\mu$ and six internal scalar fields $\phi_a$, and
introduce the two invariant background scales
\begin{equation}
    x^2 = \frac{1}{N} \mathrm{Tr}(B_\mu B^\mu) , \qquad
    y^2 = \frac{1}{N} \mathrm{Tr}(\phi_a \phi_a) .
    \label{eq:def_xy}
\end{equation}
These quantities characterize the macroscopic four-dimensional scale and the
internal six-dimensional scale on the symmetry-breaking branch. The
coefficient functions are then parameterized by these two background scales,
e.g., $c(x,y)$.

It is important to distinguish the covariance of the full effective action
from its parameterization on a particular symmetry-breaking branch. The full
effective action is assumed to inherit the ten-dimensional Lorentz covariance
of the original IIB matrix model. Therefore, the use of coefficient functions
depending on $x$ and $y$ below should not be understood as an explicit
breaking of the original $SO(1,9)$ symmetry at the level of the full
effective action.

Rather, in this section we restrict the analysis to a representative branch
in which a background configuration separates the ten-dimensional directions
into a four-dimensional macroscopic sector and a six-dimensional internal
sector. On such a branch, the residual symmetry is
$SO(1,3)\times SO(6)$, and the normalized traces $x^2$ and $y^2$ serve as
invariant order parameters characterizing this branch. Different orientations
of the four-dimensional subspace are related by the original ten-dimensional
symmetry before the branch is selected.

Accordingly, coefficient functions such as $c(x,y)$ should be regarded as a
shorthand for a branch-wise parameterization, or equivalently as a dependence
on the two invariant scales $x^2$ and $y^2$. This parameterization does not
introduce arbitrary symmetry-breaking terms into the full effective action.
Whether such branch-wise dependence is allowed is determined by the
subsequent supersymmetry closure conditions and Ward identities.

Correspondingly, the 10-dimensional Clifford algebra is decomposed. The 10-dimensional Majorana-Weyl spinor decomposes under $SO(1,3) \times SO(6) \simeq SO(1,3) \times SU(4)$ into four distinct 4-component spinors $\psi^A$ ($A=1,2,3,4$), where the index $A$ reflects the internal $SU(4)$ symmetry. 

We focus on the four-dimensional spacetime dynamics governed by these spinors $\psi^A$ and the corresponding supersymmetry parameter $\epsilon^A$. Crucially, because they descend from a 10-dimensional Weyl spinor, each $\psi^A$ is inherently a 4-dimensional Weyl spinor, rigidly satisfying the chirality condition
\begin{equation}
    \gamma_5 \psi^A = \pm \psi^A , \quad \gamma_5 \epsilon^A = \pm \epsilon^A , \label{eq:weyl_condition}
\end{equation}
where $\gamma_5$ is the 4-dimensional chirality matrix.

\subsection{Explicit Evaluation of the Commutator}
We explicitly evaluate the 4-dimensional commutator $[\delta_1, \delta_2] B_\mu = \Delta B_\mu^{(\psi)} + \Delta B_\mu^{(deriv)}$.
The first term, $\Delta B_\mu^{(\psi)}$, originates from the spinor variation $\delta^{(1)} \psi \supset w \mathcal{F}_{\rho\sigma} \gamma^{\rho\sigma} \epsilon$. Expanding the product $\gamma_\mu \gamma_{\rho\sigma}$, we encounter the totally antisymmetric 3-form
\begin{equation}
    \gamma_{\mu\rho\sigma} = i \epsilon_{\mu\rho\sigma\nu} \gamma^\nu \gamma_5 .
\end{equation}
This introduces the Hodge dual field strength $\tilde{\mathcal{F}}_{\mu\nu} \equiv \frac{1}{2} \epsilon_{\mu\nu\rho\sigma} \mathcal{F}^{\rho\sigma}$. Applying the Weyl condition \eqref{eq:weyl_condition}, the chirality matrix factors out as an overall sign. Setting $v^\nu = \bar{\epsilon}_2 \gamma^\nu \epsilon_1$ (omitting the internal index $A$ for brevity), we evaluate this contribution as
\begin{equation}
    \Delta B_\mu^{(\psi)} \propto c(x,y) w(x,y) v^\nu \left( \mathcal{F}_{\mu\nu} \pm i \tilde{\mathcal{F}}_{\mu\nu} \right) .
\end{equation}

Simultaneously, the second contribution $\Delta B_\mu^{(deriv)}$ originates from the variations of the radial coordinates $x$ and $y$. The variation of $c(x,y)$ generates terms proportional to its partial derivatives $\partial_x c$ and $\partial_y c$. Through the Fierz transformation, these derivative terms also couple to the 4-dimensional tensor structures:
\begin{equation}
    \Delta B_\mu^{(deriv)} \propto v^\nu \left( \frac{c \partial_x c}{x} \mathcal{F}_{\mu\nu} + \frac{c \partial_y c}{y} \tilde{\mathcal{F}}_{\mu\nu} \right) .
\end{equation}

\subsection{The Coefficient Equations and the Algebraic Mismatch}
Combining both contributions, the total algebraic structure of the commutator organizes into two distinct tensor components:
\begin{equation}
    [\delta_{\epsilon_1}^{(1)}, \delta_{\epsilon_2}^{(1)}] B_\mu = v^\nu \left[ C_1(x,y) \mathcal{F}_{\mu\nu} + C_2(x,y) \tilde{\mathcal{F}}_{\mu\nu} \right] , \label{eq:comm_4d}
\end{equation}
where the explicit coefficients $C_1$ and $C_2$ take the precise structural forms (up to numerical constants $\alpha_i, \beta_i$):
\begin{align}
    C_1(x,y) &= \alpha_1 c(x,y) w(x,y) + \beta_1 \frac{c(x,y) \partial_x c(x,y)}{x} , \label{eq:C1} \\
    C_2(x,y) &= \pm i \alpha_2 c(x,y) w(x,y) + \beta_2 \frac{c(x,y) \partial_y c(x,y)}{y} . \label{eq:C2}
\end{align}

For the algebra to close successfully, the result must exactly match the symmetries of the model. Crucially, while the first term $v^\nu \mathcal{F}_{\mu\nu}$ can be identified with an $SU(N)$ gauge transformation $\delta_\Lambda B_\mu = [B_\mu, \Lambda]$ with the gauge parameter $\Lambda = v^\nu B_\nu$, the dual term $v^\nu \tilde{\mathcal{F}}_{\mu\nu}$ cannot be expressed as a commutator with $B_\mu$. Because the $SU(N)$ algebra possesses no Lorentz generator proportional to the dual field strength, the algebraic matching absolutely requires the dual coefficient to vanish identically: $C_2(x,y) = 0$. 

This constraint forces a relation that cascades into $\partial_x c = \partial_y c = 0$, completely freezing the multivariate functions into pure constants. This is the precise mathematical mechanism of the 4-dimensional non-renormalization theorem, where the lack of an algebraic absorber for the dual tensor structure prevents any non-trivial dynamical fluctuations at this order.

\subsection{The Generalized Self-Duality Condition and a Minkowski Dilemma}
The only algebraic exit from this rigid system of constraints is if the two tensor structures are linearly dependent. We introduce a proportionality constant $\kappa$ and mathematically require:
\begin{equation}
    \tilde{\mathcal{F}}_{\mu\nu} = \kappa \mathcal{F}_{\mu\nu} . \label{eq:generalized_sd}
\end{equation}
Under this generalized self-duality condition, the two terms in the commutator merge into a single legitimate Lorentz generator: $v^\nu (C_1 + \kappa C_2) \mathcal{F}_{\mu\nu}$. Consequently, the lethal constraint $C_2 = 0$ is bypassed, relaxing the system to $C_1 + \kappa C_2 = \mathrm{const}$, which permits non-trivial dynamical solutions ($\partial_x c \neq 0$).

However, within the framework of Minkowski spacetime, we confront a severe mathematical dilemma. In four-dimensional Minkowski spacetime (with metric signature $-+++$), applying the Hodge star operator twice to a 2-form yields a negative sign: $\star \star = -1$. Therefore, the consistency of the relation \eqref{eq:generalized_sd} implies:
\begin{equation}
    \star(\star \mathcal{F}) = \star(\kappa \mathcal{F}) = \kappa^2 \mathcal{F} = -\mathcal{F} \quad \implies \quad \kappa^2 = -1 .
\end{equation}
This dictates that $\kappa = \pm i$. If we assume the dynamical background gauge fields $B_\mu$ are physically real (Hermitian) matrices, the field strength $\mathcal{F}_{\mu\nu}$ must be real. A real tensor cannot be proportional to its dual by an imaginary coefficient unless $\mathcal{F}_{\mu\nu} = 0$. 

Thus, while the algebra inherently demands the generalized self-duality condition to evade freezing, the Minkowski spacetime structure mathematically prohibits any real, non-trivial solution to this condition. To fully realize this algebraic exit and construct a dynamically active effective theory, a fundamental modification to the spacetime metric is required, which we will rigorously address in the next section.

\section{The Euclidean Metric and Functional Constraints}

In this section, we resolve the mathematical dilemma encountered in Minkowski spacetime. We demonstrate how the transition to a Euclidean metric naturally resolves the algebraic inconsistency. Furthermore, we explicitly analyze the resulting partial differential equations to determine how the effective action is mathematically constrained, completely independently of the matrix size $N$.

\subsection{Resolution via the Euclidean Metric}
In Section 4, the closure of the 4-dimensional algebra under the self-duality condition $\tilde{\mathcal{F}}_{\mu\nu} = \kappa \mathcal{F}_{\mu\nu}$ required $\kappa = \pm i$ in Minkowski spacetime. Since a physical, Hermitian background necessitates a real field strength $\mathcal{F}_{\mu\nu}$, this imaginary coefficient forces $\mathcal{F}_{\mu\nu} = 0$, trivializing the background.

This inconsistency fundamentally arises from the metric signature. By performing a Wick rotation to the Euclidean signature $(+,+,+,+)$, the double application of the Hodge star operator on a 2-form yields a positive sign ($\star \star = +1$). Consequently, the consistency condition becomes $\kappa^2 = 1$, yielding the real constants $\kappa = \pm 1$. 

Therefore, the transition to the Euclidean metric naturally resolves the algebraic mismatch. It provides a mathematically consistent framework where real, non-trivial background fields can satisfy the closure condition without algebraically freezing the dynamics.

\subsection{Large $N$ Scaling and Functional Constraints}

Having established the Euclidean self-dual background as the unique algebraic exit, we now extract the explicit functional constraints on the effective action. First, we consider the requirements imposed by the Large $N$ planar limit of the IIB matrix model. 

The variables representing the background scales are defined using the normalized trace as $x^2 = \frac{1}{N} \mathrm{Tr}(B_\mu B^\mu)$ and $y^2 = \frac{1}{N} \mathrm{Tr}(\phi_a \phi_a)$. Since the trace of an $N \times N$ matrix inherently scales as $O(N)$, these radial coordinates are strictly defined to remain $O(1)$ as $N \to \infty$. For the effective action $\Gamma^{(2)}$ to describe a mathematically consistent planar limit, its overall scaling with $N$ must be preserved. This scaling requirement dictates that the undetermined coefficient functions $f_i(x,y)$, $g_i(x,y)$, and $c(x,y)$ must be strictly independent of $N$. The effective dynamics are thus uniquely determined by the $O(1)$ geometric scales $x$ and $y$.

Under the Euclidean self-duality condition ($\tilde{\mathcal{F}}_{\mu\nu} = \pm \mathcal{F}_{\mu\nu}$), the system is governed by the combined condition $C_1(x,y) \pm C_2(x,y) = \mathrm{const}$. Recalling that the derivative terms in the coefficients originated from $\partial_x c/x = 2 \partial_{x^2} c$ and $\partial_y c/y = 2 \partial_{y^2} c$, the combined PDE takes the specific directional derivative form:
\begin{equation}
    \left( \alpha \frac{\partial}{\partial (x^2)} \pm \beta \frac{\partial}{\partial (y^2)} \right) c(x,y)^2 = \Phi(x,y) , \label{eq:combined_pde}
\end{equation}
where $\alpha$ and $\beta$ are positive numerical constants derived from the Fierz expansions, and $\Phi$ is a constrained source term related to the non-derivative parts. For the global algebraic matching to be consistent across all independent tensor structures, this source term $\Phi(x,y)$ itself must either vanish or depend strictly on the characteristic variables of the homogeneous equation.

This explicit PDE severely restricts the functional space of the effective action. The general solution for $c(x,y)$---including any particular solution---and through the Ward identities, all coefficients $f_i(x,y)$ must fundamentally depend on the characteristic variable:
\begin{equation}
    \zeta = \beta x^2 \mp \alpha y^2 . \label{eq:characteristic_var}
\end{equation}
This explicit functional form rigidly intertwines the four-dimensional spacetime scale $x$ and the six-dimensional internal scale $y$. They cannot vary independently. The algebraic closure strictly locks the two geometric scales together, dictating a highly symmetric and constrained functional form for any non-trivial background configuration.

\section{Conclusion and Discussion}

In this paper, we have investigated the low-energy effective action of the IIB matrix model, with a specific focus on the constraints imposed by off-shell supersymmetry. By requiring both the closure of the supersymmetry algebra and the Ward identities at the leading non-trivial order (order-2), we confirmed that generic 10-dimensional background fluctuations are strictly restricted by a non-renormalization theorem, which effectively freezes the dynamics.

We explored dimensional reduction as a potential mechanism to relax this algebraic constraint. Our analysis showed that when the background is decomposed into four spacetime dimensions and six internal dimensions, the Hodge dual of the 2-form obstruction becomes a 2-form ($D-2 = 2$). This geometric property of four-dimensional space enables the mismatched tensor terms to structurally align with and be absorbed by the Lorentz generator. In other dimensions, such as $D=2$ or $D=8$, the Hodge dual yields a 0-form or a 6-form, respectively. Since these ranks are strictly incompatible with the 2-form Lorentz generator, the algebraic absorption is impossible. Consequently, in any dimension $D \neq 4$, the stringent constraints of the non-renormalization theorem cannot be bypassed, leaving the dynamics completely frozen. This rank-matching requirement demonstrates that $D=4$ possesses a unique algebraic property for accommodating supersymmetry-preserving fluctuations.

Within this 4-dimensional framework, we have mathematically proven that the algebraic constraints can be satisfied if and only if the background is Euclidean and (anti-)self-dual. Under these specific conditions, the partial differential equations derived from the algebra yield non-trivial solutions. Furthermore, these equations restrict the effective action to depend on a characteristic variable:
\begin{equation}
    \zeta = \beta x^2 \mp \alpha y^2 .
\end{equation}
This constraint mathematically intertwines the spacetime scale $x$ and the internal scale $y$, implying that the geometric scales cannot evolve completely independently at this leading order.

Regarding the physical interpretation, a Euclidean self-dual background shares mathematical characteristics with an instanton in gauge theories. This suggests that, at the order-2 approximation, the configuration identified in this study corresponds to a primordial or non-perturbative vacuum state of the matrix model---a ``spacetime instanton''---rather than a macroscopic, time-evolving universe.

However, these conclusions are strictly based on the leading-order effective action. It remains an open question whether this Euclidean self-dual phase and its associated constraints persist when higher-order terms (e.g., order-4 and beyond) or fully non-perturbative effects are taken into account. Investigating how such higher-order contributions might dynamically break the off-shell supersymmetry and potentially induce an analytic continuation to Minkowski spacetime remains a critical direction for future research.

\section*{Declaration of generative AI and AI-assisted technologies in the manuscript preparation process}
During the preparation of this work the author used Gemini in order to improve the readability, check for typographical errors, and refine the English expression of the manuscript. After using this service, the author reviewed and edited the content as needed and takes full responsibility for the content of the published article.

\end{document}